# Decoding the Digital Fine Print: Navigating the potholes in Terms of service/ use of GenAI tools against the emerging need for Transparent and Trustworthy Tech Futures

Sundaraparipurnan Narayanan

Doctoral Candidate, University of Bordeaux, sundar.narayanan@aitechethics.com

The research investigates the crucial role of clear and intelligible terms of service in cultivating user trust and facilitating informed decision-making in the context of AI, in specific GenAI. It highlights the obstacles presented by complex legal terminology and detailed fine print, which impede genuine user consent and recourse, particularly during instances of algorithmic malfunctions, hazards, damages, or inequities, while stressing the necessity of employing machine-readable terms for effective service licensing. The increasing reliance on General Artificial Intelligence (GenAI) tools necessitates transparent, comprehensible, and standardized terms of use, which facilitate informed decision-making while fostering trust among stakeholders. Despite recent efforts promoting transparency via system and model cards, existing documentation frequently falls short of providing adequate disclosures, leaving users ill-equipped to evaluate potential risks and harms. To address this gap, this research examines key considerations necessary in terms of use or terms of service for Generative AI tools, drawing insights from multiple studies. Subsequently, this research evaluates whether the terms of use or terms of service of prominent Generative AI tools against the identified considerations. Findings indicate inconsistencies and variability in document quality, signaling a pressing demand for uniformity in disclosure practices. Consequently, this study advocates for robust, enforceable standards ensuring complete and intelligible disclosures prior to the release of GenAI tools, thereby empowering end-users to make well-informed choices and enhancing overall accountability in the field.

## 1 INTRODUCTION

Terms of service (ToS) are crucial legal documents that define the rules of conduct for customers and guarantee the availability of services and resources when the conditions are met [1]. Privacy policies, a key component of ToS, play a significant role in influencing users' perceived effectiveness of privacy policy, with transparency, control, and protection being important factors [2]. However, research indicates that users rarely read privacy policies, despite claiming to be very concerned about their privacy [3]. This lack of attention to privacy policies is attributed to their length, legal jargon, and the uncertainty of whether the service behaves as described in the policy [4]. [5] and [6] [7] both emphasize the importance of informed consent and transparency in AI systems. [5] highlights the need for users to understand how these systems operate, while [7] proposes the use of Data Cards to provide clear and comprehensive information about the datasets used. These measures can help mitigate bias and build trust, as suggested by [8] and [9]. [8] stresses the role of policy and governance in ensuring responsible AI use, while [9] suggests incorporating lessons from information security to enhance the development of AI systems. Prior research thereby clarifies the need for transparency, accountability, and user empowerment in the design and deployment of AI technologies.

## 1.1 Emergence of model cards, system cards, and data cards

The Model Card initiative by Google and the Datasheets for Datasets concept have emerged as essential tools for promoting transparency and accountability in machine learning systems. Model Cards provide a structured format for reporting on model performance and potential biases, enhancing external user-readiness with machine learning systems [10]. On the other hand, Datasheets for Datasets aim to facilitate communication between dataset creators and consumers, prioritizing transparency and accountability in the machine learning community [11]. These initiatives reflect a growing emphasis on documentation and transparency in the development and deployment of AI-enabled systems, aligning with the broader goal of promoting responsible and ethical AI practices [12].

The incorporation of algorithmic model cards, system cards, and datasheets for datasets is crucial for ensuring transparency, accountability, governance, and respect for user rights [13] [6] [7] [14] [15] [16]. These tools provide detailed information about the datasets development, intent, and ethical considerations, as well as the processes and rationales that shape the data and models [6] [7]. They also offer guidance on mitigating potential shortcomings and improving system performance [14]. Furthermore, they support algorithmic accountability by increasing transparency, monitoring outcomes, and improving governance [15] In the context of public policy, system cards can serve as scorecards for algorithm audits, ensuring that AI-based decision-aiding systems meet appropriate accountability standards [16].

## 1.2 Emergence of LLMs

The emergence of Large Language Models (LLMs) has revolutionized the field of natural language processing, leading to significant advancements in various downstream tasks. Notable LLMs such as GPT4 [17], LLama2 [18], Falcon [19], and Mistral [20] have garnered attention due to their adaptability and effectiveness across diverse applications. Research focusing on open-sourced LLMs [21] has democratized advanced language models, enabling widespread access and utilization.

These LLMs have driven innovations in several key areas, including adapting LLMs, prompt engineering, emergent abilities, and optimization strategies. In the area of adaption of LLMs, advanced techniques such as leveraging large context length [22] [23] [24], and fine-tuning methods like instruction tuning [25], alignment tuning, & memory-efficient model adaptation [26] [27] [28] have emerged. These approaches allow LLMs to be fine-tuned for specific tasks and domains, enhancing their performance and applicability to targeted applications.

## 2 TERMS OF SERVICE AND ITS CONTENTS

Terms of service, also known as terms of use, encompass a range of clauses and legal provisions that outline the rules and conditions governing the use of a service or product. These typically include a license grant for using the service, separate terms for open-source components if applicable, and rules regarding confidentiality, privacy, security, and safety. Restrictions on tool use, remedies for breaches, and the availability of the service are also commonly covered.

Additionally, terms of service may address the right to remove content that violates third-party rights, confidentiality terms, privacy and data protection provisions, security measures for using the tool, safety considerations, and the use of customer data for training and monitoring purposes, intellectual property rights, warranty information, and policies related to service outages, interruptions, and changes. These terms serve to protect both the service provider and the user, outlining their respective rights and responsibilities, and they may also specify details about disputes, jurisdiction, and conditions for termination providing legal terminology.

They may not always include sections on unacceptable uses of the service, the feedback process, and technical support contact information. In addition, aspects such as the intended use of the service, model information, training data, performance metrics, bias metrics, and safety metrics are often not covered comprehensively in traditional terms of service agreements. Despite recent efforts promoting transparency via system and model cards, existing documentation frequently falls short of providing adequate disclosures, leaving users ill-equipped to evaluate potential risks and harms. These aspects are useful particularly during instances of algorithmic malfunctions, hazards, damage, or inequities, impeding genuine user consent and recourse.

This research investigates the crucial role of clear and intelligible terms of service in cultivating user trust and facilitating informed decision-making in the context of AI, in specific Generative AI.

## 3 EXPERIMENT DESIGN

Given the democratization of the use of Generative AI tools, the paper explores the nature, extent, and approach towards providing sufficient information to the user, stressing the necessity of employing machine-readable terms for effective service licensing. To that end, 86 Generative Ai companies belonging to 13 categories were identified and their respective terms were analyzed.

### 3.1 Methodology of examining the terms of service

Firstly, the research identified a diverse range of Generative AI companies spanning various domains. Secondly, the websites of these companies were visited to collect the most recent versions of their terms of service documents. Thirdly, specific aspects, including intended use, model and data information, limitations, open-source terms, were determined for evaluation. Fourthly, each company's terms of service were reviewed to identify gaps, inconsistencies, and the presence or absence of the determined aspects, with a focus on user-centric principles. Finally, the findings were then compiled into a structured dataset and were analyzed to qualitatively gauge the trends arising from the evaluation.

### 3.2 Identification of Generative AI companies

Dealroom.co lists 446 Generative AI companies along with details of number of employees, extent of funding, revenue, market, focus and location details. Of the above list, 86 generative AI companies belonging to 13 categories with over 50 employees were identified for the research. The research considered 50 employees to be a representative number, reflecting the growth and capacity of the organization to be able to meaningfully contribute to better terms of service.

The 13 categories the companies belonged include text (copy & writing, customer relations, knowledge & research, and note taker assistants), audio (covering speech generation and music generation), video, synthetic data, code generation or repository, gaming (covering performance analytics, 3d design and game moderation), image (covering image generation and design & marketing), legal, LLM tools (data and embedding management, deployment and monitoring, performance management, model development and fine-tuning), personal Ai assistant, cloud compute providers, general intelligence and model creators and others (including art generators, character creation). Funding for these companies varied between USD 3mn to USD 1.4bn

## 4 EVALUATION RESULTS

### 4.1 Trends from the evaluation

Based on the evaluation, the following trends were observed:

*4.1.1 Missing terms of service*
22 out of the 86 companies did not have any terms of service or terms of use on their webpage. Out of the 22, 6 of them had privacy policies, 1 of them had website terms and the rest did not have any terms on their webpage.

*4.1.2 Recency of updating of terms*
Of the 64 companies, 17 of them were undated. 16 companies had updated their terms prior to 2023, despite the extent of changes that emerged in the year 2023 for Generative AI. And the rest of them were updated during the year 2023 & 2024.

*4.1.3 Intended use*

33 out of 64 companies did not express the intended use of the tool. The intend use of the product typically provide information about how the services should be used, including details on the platform's functionalities, purposes, and compliance with relevant laws and policies. Some descriptions highlighted specific use cases, such as processing health data, content creation, social connectivity, or providing a cloud-based proprietary platform.

*4.1.4 Model cards, data cards or system cards*

None of the 63 companies (excluding huggingface) mentioned about the model they used, the training data , the performance metrics, bias metrics, safety and security metrics or any form of model card or system card or any reference to such a document as part of the terms of use or terms of service.

*4.1.5 Limitations of the tool*

51 out of 64 companies did not provide any references as to the limitations of the tool as part of the terms of use or terms of service. Some sources referred to the limitations of tools, solutions, or services, while others mention restrictions on the use of specific platforms or databases. However, do not explicitly focus on the risks, errors, or constraints associated with the respective tools or services in specific.

*4.1.6 Customer data management*

44 out of 64 companies do not mention anything about how they deal with customer data and whether they use them to train their model, improve efficiency or for monitoring.

*4.1.7 Unacceptable uses of the tool*

36 out of 64 companies do have specific clauses that explain the unacceptable uses of the tool. It is pertinent to note that fake news, deep fake and misinformation is increasing with Generative AI tools.

*4.1.8 Limitations to reverse engineering of the tool*

45 out of the 64 did not restrict or prohibit reverse engineering the tool.

*4.1.9 Feedback process and support contact*

54 out of 64 companies do not mention feedback process or feedback mechanism for the user to interact with to provide the feedback. 55 of the 64 did not have any support contacts for the user to reach out to.

*4.1.10 Intellectual property*

15 out of 64 companies did not provide the details of intellectual property rights as part of the terms. Of the rest, one of them (Anyword) claimed IP for content generated on the platform.

*4.1.11 Warranty*

All of the companies disclaimed the warranty as part of their terms of use or terms of service.

*4.1.12 Outage, service interruption, or changes to the service*

50 out of 64 did not have any terms regarding outage of service, service interruptions, changes to the service.

*4.1.13 Limitations of liability*

60 of them mentioned clauses regarding limitations of liability. 39 of the 64 mentioned confidentiality, 54 mentions about privacy and data protection, but does not adequately cover safety and security & indemnity.

## 4.2 Discussion and way forward

The utilization of Generative AI, including large language models (LLMs), in downstream applications carries various risks, including privacy concerns. These models can capture sensitive information and may lead to various ethical and social risks, including discrimination, misinformation, and misuse [29] [30] [31].

Incorporating specific terms in the terms of use or terms of service agreements is essential for ensuring transparency, user understanding, and future legal compliance. Expressing a product's intended use provides clear usage guidelines while disclosing model and data information promotes transparency and risk assessment. Describing limitations helps users set realistic expectations, and addressing open-source terms clarifies licensing conditions. Explaining how customer data is handled safeguards privacy, and defining unacceptable uses sets ethical boundaries. Restrictions on reverse engineering protect intellectual property, and providing feedback mechanisms and support contacts fosters user communication and service improvement. Outlining intellectual property rights defines content ownership while disclaiming warranties and addressing service-related terms to manage user expectations and legal liabilities.

The above findings indicate inconsistencies and variability in terms' quality, requiring standards demanding uniformity in disclosure practices. Consequently, this study advocates for robust, enforceable standards ensuring complete and intelligible disclosures before the release of Generative AI tools, thereby empowering end-users to make well-informed choices and enhancing overall accountability in the field. This paper contributes to the ongoing discourse on responsible AI development by providing insights into the current state of terms of use in the AI domain and advocating for robust standards that can contribute to a more transparent and accountable AI ecosystem.

## ACKNOWLEDGMENTS

This project is not supported by any grants.

## 5 HISTORY DATES

The submission is not currently presented to any journals.